\documentclass[aip,jap,preprint]{revtex4-1}
\pdfoutput=1
\usepackage[pdftex]{graphicx}
\usepackage{dcolumn}
\usepackage{bm}
\usepackage{amsmath}

\begin{document}

\title{Finite-displacement elastic solution due to a triple contact line}

\affiliation{Postal address: Juan Olives Ba\~nos, Poste restante, 215 boulevard Paul Claudel, 13010 Marseille, France}

\author{Juan Olives}

\email{olivesbanosjuan@gmail.com.\\ORCID 0000-0002-0087-3165.\\Before at CINaM-CNRS Aix-Marseille Univ., Campus de Luminy, 13288 Marseille Cedex 9, France.}

\date{\today}

\begin{abstract}
At the line of triple contact of an elastic body with two immiscible fluids, the body is subjected to a force concentrated on this line, the fluid--fluid surface tension. In the simple case of a semi-infinite body, limited by a plane, a straight contact line on this plane, and a fluid--fluid surface tension normal to the plane, the classical elastic solution leads to an infinite displacement at the contact line and an infinite elastic energy. By taking into account the body--fluid surface tension (i.e., isotropic surface stress), we present a new and more realistic solution concerning the semi-infinite body, which gives a finite displacement and a ridge at the contact line, and a finite elastic energy. This solution also shows that Green's formulae, in the volume and on the surfaces, are valid (these formulae play a central role in the theory).
\end{abstract}

\maketitle

\section{Introduction}

Surface properties of deformable bodies have many applications, e.g., in adhesion, coating, thin films and nanosciences. When a deformable body is in contact with two fluids (e.g., liquid and air), it is subjected to the fluid--fluid surface tension along the body--fluid--fluid contact line. The classical solution for the deformation of an elastic body, occupying a semi-infinite space limited by a plane, and subjected to a fluid--fluid surface tension concentrated on a straight line of this plane and normal to the plane, leads to an infinite displacement at this line and an infinite elastic energy.\cite{Flamant:1892} However, the presence of a ridge at the contact line was experimentally observed\cite{Pericet-Camara-etal:2009} and then confirmed by other experiments and Fourier transform calculations.\cite{Jerison-etal:2011} In our previous work,\cite{Olives:2010a} by taking into account the body--fluid surface stresses and surface energies, we showed that there are two equilibrium equations at the contact line: (i) the equilibrium of the three surface stresses (with no contribution of the volume stresses); (ii) a scalar equation involving the surface energies, the surface stresses, and the surface strains, which leads to a generalization of the classical Young's equation (only valid for a rigid body) (see Sec.~\ref{GEN}). The first equation---as an equilibrium of three forces tangent to the three interfaces---implies a finite displacement and the formation of a ridge at the contact line. This equation was then experimentally confirmed.\cite{Style-etal:2013} In the present paper, by taking into account the body--fluid surface tension (i.e., isotropic surface stress), we obtain a new solution of the above elastic problem concerning the semi-infinite body, with a finite displacement and a ridge at the contact line, and a finite elastic energy.

\section{General equilibrium equations} \label{GEN}

In the previous work Ref.~\onlinecite{Olives:2010a} (with additional comments in Ref.~\onlinecite{Olives:2010b} and mathematical aspects in Ref.~\onlinecite{Olives:2014}), following the general variational method of Gibbs,\cite{Gibbs:1878} we gave the thermodynamic definition and properties of the surface stress and obtained the following equilibrium equations {\it for any deformable body}:

1) On each body--fluid surface:
\begin{align}
&{\rm div}\,\bar{\sigma_{\rm s}} + \rho_{\rm s}\,\bar g + \sigma \cdot n + p\,n
= 0 \label{bfequilibrium}\\
&\sigma_{\rm s}^* = \sigma_{\rm s},
\end{align}
where $\sigma_{\rm s}$ is the body--fluid surface stress, $\rho_{\rm s}$ the surface mass excess per unit area, $\bar g$ the gravity field, $\sigma$ the body volume stress, $n$ the unit vector normal to the surface (oriented from the fluid to the body), and $p$ the fluid pressure; $\bar{\sigma_{\rm s}}=\iota \cdot \sigma_{\rm s}$, where $\iota$ is the natural injection of the tangent plane to the surface ${\rm T}_x(\rm S)$ in the three-dimensional space $\rm E$ (i.e., $\bar{\sigma_{\rm s}}^{\beta i}=\sigma_{\rm s}^{\alpha\beta}\,\partial_\alpha x^i$, with components $\alpha$ and $\beta$ on the surface, and $i$ in $\rm E$), and ${\rm div}\,\bar{\sigma_{\rm s}}$ is a special divergence based on the tensorial product of the covariant derivative on the surface and the usual derivative in $\rm E$, defined in Ref.~\onlinecite{Olives:2014} (i.e., with components: $({\rm div}\,\bar{\sigma_{\rm s}})^i=\partial_\beta (\sigma_{\rm s}^{\alpha\beta}\,\partial_\alpha x^i) + \Gamma_{\beta\gamma}^\beta\,\sigma_{\rm s}^{\alpha\gamma}\,\partial_\alpha x^i$). Eq.~(\ref{bfequilibrium}) has a tangential component
\begin{eqnarray}
{\rm div}\,\sigma_{\rm s} + \rho_{\rm s}\,{\bar g}_t + (\sigma \cdot n)_t = 0 \label{bfequilibriumtangent}
\end{eqnarray}
(in which ${\rm div}\,\sigma_{\rm s}$ is the usual surface divergence and the subscript $t$ indicates the vector component tangent to the surface) and a normal component
\begin{eqnarray}
l_n : \sigma_{\rm s} + \rho_{\rm s}\,{\bar g}_n + \sigma_{nn} + p = 0, \label{bfequilibriumnormal}
\end{eqnarray}
where $l$ is the curvature fundamental form on the surface, $l_n = l \cdot n$ (i.e., with components: $l_{\alpha\beta}^i = \partial_{\alpha\beta} x^i - \Gamma_{\alpha\beta}^\gamma\,\partial_\gamma x^i$, $l_{n,\alpha\beta} = l_{\alpha\beta}^i\,n_i$; the symbol : means double contraction of the indices), ${\bar g}_n = \bar g \cdot n$, and $\sigma_{nn} = (\sigma \cdot n) \cdot n$. The eigenvalues of $l_n$ are the principal curvatures, $\frac{1}{R_1}$ and $\frac{1}{R_2}$, of the surface (a curvature being positive when its center is on the side of $n$). If $\sigma_{\rm s}$ is isotropic, i.e., $\sigma_{\rm s} = \hat{\sigma_{\rm s}}\,I$ (eigenvalue $\hat{\sigma_{\rm s}}$ and $I$ the identity), we have
\begin{eqnarray}
l_n : \sigma_{\rm s} = \hat{\sigma_{\rm s}} (\frac{1}{R_1} + \frac{1}{R_2}). \label{curvature}
\end{eqnarray}
Similar equations were previously written, but under some particular assumptions, e.g., the existence of a ``surface traction field''\cite{Gurtin-Murdoch:1975} or in the special case of elastic bodies.\cite{Alexander-Johnson:1985, Gurtin-etal:1998}

2) On the body--fluid--fluid contact line, there are two equations (as previously found in the particular case of the elastic thin plate\cite{Olives:1993, Olives:1996}):

The first one is vectorial (three-dimensional) and corresponds to a line fixed on the body (but the line can move because the body is deformable):
\begin{eqnarray}
\sigma_{\rm bf} \cdot \nu_{\rm bf} + \sigma_{\rm bf'} \cdot \nu_{\rm bf'}
+ \gamma_{\rm ff'}\,\nu_{\rm ff'} = 0, \label{bff'equilibrium1}
\end{eqnarray}
in which the subscripts $\rm b$, $\rm f$, and $\rm f'$ respectively denote the body and the two fluids, $\sigma_{\rm bf}$ is the $\rm bf$ surface stress, $\nu_{\rm bf}$ the unit vector normal to the contact line and tangent to the $\rm bf$ surface (directed to the inside of $\rm bf$), idem for $\sigma_{\rm bf'}$, $\nu_{\rm bf'}$, and $\nu_{\rm ff'}$, and $\gamma_{\rm ff'}$ is the $\rm ff'$ surface tension. This equation expresses the equilibrium of the three surface stresses acting on the contact line (with no contribution of the volume stresses), and determines the angles of contact $\varphi_{\rm f}$, $\varphi_{\rm f'}$, and $\varphi_{\rm b}$, respectively measured in $\rm f$, $\rm f'$, and $\rm b$ (satisfying $\varphi_{\rm f} + \varphi_{\rm f'} + \varphi_{\rm b} = 2\pi$). This point was then experimentally verified in Ref.~\onlinecite{Style-etal:2013}.

The second equation is scalar and corresponds to a line moving with respect to the body, but fixed in space (i.e., the displacement of the material points of the body, due to the deformation, exactly compensates the displacement of the line with respect to the body, so that the line remains fixed in space):
\begin{align}
&(\sigma_{{\rm bf},\nu\nu} - \gamma_{\rm bf})\,a_{\nu\nu}
- (\sigma_{{\rm bf'},\nu\nu} - \gamma_{\rm bf'})\,a'_{\nu\nu}\nonumber\\
&+ \sigma_{{\rm bf'},\tau\nu}(a'_{\tau\nu} - a_{\tau\nu}) = 0, \label{bff'equilibrium2a}
\end{align}
in which $\tau$ is a unit vector tangent to the line, $\sigma_{\rm bf,\nu\nu}$ and $\sigma_{\rm bf,\tau\nu}$ are respectively the components along $\nu_{\rm bf}$ and $\tau$ of the $\rm bf$ surface stress acting on the line, idem for $\sigma_{\rm bf',\nu\nu}$ and $\sigma_{\rm bf',\tau\nu}$, $\gamma_{\rm bf}$ and $\gamma_{\rm bf'}$ are respectively the $\rm bf$ and $\rm bf'$ surface energies, $a_{\nu\nu}$ is the surface stretching deformation, normal to the line, in the $\rm bf$ side, $a_{\tau\nu}$ the surface shear deformation, parallel to the line, in the $\rm bf$ side, and idem for $a'_{\nu\nu}$ and $a'_{\tau\nu}$ in the $\rm bf'$ side. An equivalent form of this equation is (with the help of Eq.~(\ref{bff'equilibrium1}))
\begin{align} 
&-\gamma_{\rm bf} + \gamma_{\rm bf'}\,a_{\rm r,\nu\nu}- \gamma_{\rm ff'}\,\cos\varphi_{\rm f} \nonumber\\
&- \gamma_{\rm ff'}\,\sin\varphi_{\rm f}\,
\frac{\cos\varphi_{\rm b} + a_{\rm r,\nu\nu}}{\sin\varphi_{\rm b}}
+ \sigma_{\rm bf',\tau\nu}\,a_{\rm r,\tau\nu} = 0,\label{bff'equilibrium2b} 
\end{align}
where $a_{\rm r,\nu\nu} = \frac{a'_{\nu\nu}}{a_{\nu\nu}}$ and $a_{{\rm r},\tau\nu} = \frac{a'_{\tau\nu} - a_{\tau\nu}}{ a_{\nu\nu}}$. 

Note that, for a perfectly rigid body, the above Eqs.~(\ref{bff'equilibrium1}) and (\ref{bff'equilibrium2a}) cannot be written (since they are based on the possible displacement of the material points, due to the deformation) and, in this case, we obtain the unique scalar equation
\begin{align}
 -\gamma_{\rm bf} + \gamma_{\rm bf'} - \gamma_{\rm ff'}\,\cos\varphi_{\rm f} = 0, \label{Young}
\end{align}
which is the classical Young's equation (in which $\gamma_{\rm bf}$ and $\gamma_{\rm bf'}$ are surface energies, and {\it not} surface tensions). If the body is deformable, Young's equation is not valid and the valid Eq.~(\ref{bff'equilibrium2b}) is the generalization of Young's equation. Indeed, in the rigid body limit, $a_{\rm r,\nu\nu} = 1$, $a_{\rm r,\tau\nu} = 0$, and $\varphi_{\rm b} = \pi$ (thus, $\lim_{\varphi_{\rm b} \rightarrow \pi} \frac{\cos\varphi_{\rm b} + 1}{\sin\varphi_{\rm b}} = 0$), and Eq.~(\ref{bff'equilibrium2b}) leads to Young's equation. Also note that, in the case of a very little deformable body, i.e., an almost rigid body, the surface (and volume) stresses become almost infinite, so that the two first terms in Eq.~(\ref{bff'equilibrium1}) are almost infinite and this equation implies that $\varphi_{\rm b}$ is almost equal to $\pi$ (in order to equilibrate the finite fluid-fluid surface tension), i.e., that there is almost no ridge on the surface. Finally, note that, if the body is a fluid, then $\sigma_{\rm bf} - \gamma_{\rm bf}\,I = \sigma_{\rm bf'} - \gamma_{\rm bf'}\,I = 0$ (which implies that $\sigma_{{\rm bf},\nu\nu} - \gamma_{\rm bf} = \sigma_{{\rm bf'},\nu\nu} - \gamma_{\rm bf'} = \sigma_{{\rm bf'},\tau\nu} = 0$) so that Eq.~(\ref{bff'equilibrium2a}) is useless, and Eq.~(\ref{bff'equilibrium1}) expresses the equilibrium of the three fluid--fluid surface tensions.

Eq.~(\ref{bff'equilibrium2a}) expresses that the variation of surface energy (e.g., increase in ${\rm bf}$ surface and decrease in ${\rm bf'}$ surface, when the line moves with respect to the body) is equal to the work of the surface stresses acting on the line (due to the displacement of the material points, on each side of the line), when the line remains fixed in space. This equation gives a condition on the components $a_{\rm r,\nu\nu}$ and $a_{\rm r,\tau\nu}$ of the ``relative'' surface deformation of the $\rm bf'$ side with respect to the $\rm bf$ side (we defined the ``relative deformation gradient'' in Ref.~\onlinecite{Olives-Bronner:1984}). If $\sigma_{{\rm bf'},\tau\nu} = 0$ (which, e.g., occurs if the surface stress is isotropic), it gives the relative surface stretching deformation $a_{\rm r,\nu\nu}$ (normal to the line) of the $\rm bf'$ side with respect to the $\rm bf$ side.

\section{Validity of Green's formula:\\no contribution of the volume stresses} \label{GREENval}

The preceding theory is based on Green's formula
\begin{align}
\int_{\rm V} \sigma : {\rm D}w \,dv = -\int_{\rm V} {\rm div} \sigma \cdot w \,dv - \int_{\rm S} (\sigma \cdot n) \cdot w \,da,\label{Green}
\end{align}
where $\sigma$ is the volume stress tensor, $w$ a virtual displacement, $\rm V$ a volume of the body in the neighborhood of the contact line, $\rm S$ the boundary surface, $n$ the unit vector normal to the surface and directed towards the interior of the body, $dv$ an element of volume, and $da$ an element of area. With the help of a first example of solution, we showed in Ref.~\onlinecite{Olives:2014} that:

1) Owing to the singularity at the contact line, the components of $\sigma$ do not belong to the Sobolev space $H^1(\rm V)$.

2) Nevertheless, Green's formula remains valid, because all the components $\partial_j u_i$ and $\sigma_{ij}$ are either bounded or subjected to the inequality
\begin{eqnarray}
|\partial_j u_i|, \,|\sigma_{ij}| \leq c |\log r| + d\label{inequality}
\end{eqnarray}
in $\rm V$ ($r$ is the distance to the contact line; $c$ and $d$ being positive constants), and these inequalities imply that
\begin{eqnarray}
\lim_{\varepsilon \rightarrow 0} \int_{{\rm S}(\varepsilon)} (\sigma \cdot n) \cdot w \,da = 0,\label{linevolumestress}
\end{eqnarray}
where ${\rm S}(\varepsilon)$ is the boundary surface of a small tubular volume ${\rm V}(\varepsilon)$ of the body, of radius $\varepsilon$, around an element of contact line (${\rm V}(\varepsilon)$ is bounded by (i) the surface of the body and (ii) a half-cylinder of radius $\varepsilon$ around the contact line). The validity of Green's formula is based on Eq.~(\ref{linevolumestress}), which directly expresses that the volume stresses have no contribution at the contact line.

3) The elastic energy is finite (in the neighborhood of the contact line).

These results will be confirmed with the solution given in the present paper.
 
\section{Application to the semi-infinite elastic body} \label{Application}

\begin{figure}[htbp]
\begin{center}
\includegraphics[width=16cm]{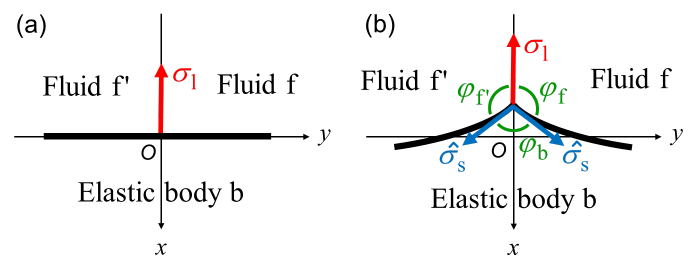}
\end{center}
\caption{Semi-infinite elastic body $\rm b$ in contact with two fluids $\rm f$ and $\rm f'$, and subjected to the fluid--fluid surface tension $\sigma_{\rm l}$ normal to the surface $x = 0$ of the body and concentrated on the line $x = y = 0$ (a). At the surface of the body, there is a constant surface tension $\hat{\sigma_{\rm s}}$. (a)~Before deformation; (b)~after deformation.} \label{Notations}
\end{figure}

Let us consider a semi-infinite isotropic elastic body $\rm b$, occupying the half space $x \geq 0$ in the orthonormal frame $(Ox, Oy, Oz')$, in contact with a fluid $\rm f$ occupying the region $x < 0$ and $y > 0$, and another fluid $\rm f'$ in the region $x < 0$ and $y < 0$. It is subjected to the fluid--fluid surface tension $\gamma_{\rm ff'}$, here denoted $\sigma_{\rm l}$, which is a force parallel to $Ox$ and concentrated on the line $x = y = 0$ (Fig.~\ref{Notations}a).
Sign convention: $\sigma_{\rm l} > 0$ if the direction of the force is opposite to $Ox$ (which is the case of a fluid--fluid surface tension), and $\sigma_{\rm l} < 0$ if the direction of the force is $Ox$ (in this case, it is a compression). We suppose that the $\rm bf$ and $\rm bf'$ surface stress tensors are isotropic, with the same constant eigenvalue, noted $\hat{\sigma_{\rm s}} > 0$ (surface tension), and that the $\rm bf$ and $\rm bf'$ surface energies are equal. In this case, Eq.~(\ref{bff'equilibrium2a}) only implies that $a_{\rm r,\nu\nu} = 1$ (same surface stretching in the $\rm bf$ side and in the $\rm bf'$ side), which will be obviously satisfied owing to the symmetry of the problem with respect to the plane $y = 0$.\cite{Note} The other Eq.~(\ref{bff'equilibrium1}) may be written as
\begin{eqnarray*}
\sigma_{\rm l} = 2 \hat{\sigma_{\rm s}} \cos\varphi,
\end{eqnarray*}
where $\varphi = \varphi_{\rm b}/2$ (see Fig.~\ref{Notations}b). Clearly, the elastic displacement components $u_x$ and $u_y$ are only functions of $(x,y)$, $u_{z'} = 0$, and, by symmetry, $u_x(x,-y) = u_x(x,y)$ and $u_y(x,-y) = -u_y(x,y)$. In the approximation of small deformations, i.e., when the components of the displacement and their first derivatives are small, we have $\cos\varphi \approx \partial_y u_x(0,0+)$ (derivative at $x = 0$, $y = 0+)$, and the preceding equation gives
\begin{eqnarray}
\sigma_{\rm l} = 2 \hat{\sigma_{\rm s}}\, \partial_y u_x(0,0+). \label{equline}
\end{eqnarray}

The Eqs.~(\ref{bfequilibriumtangent})-(\ref{bfequilibriumnormal}) on the surface $x = 0$ are (in the absence of gravity and using Eq.~(\ref{curvature}))
\begin{align*}
&(\sigma \cdot n)_t = 0,\\
&\hat{\sigma_{\rm s}} (\frac{1}{R_1} + \frac{1}{R_2}) + \sigma_{nn} + p = 0
\end{align*}
(${\rm div}\,\sigma_{\rm s} = 0$, because $\sigma_{\rm s} = \hat{\sigma_{\rm s}}\,I$ and $\hat{\sigma_{\rm s}}$ is constant on the surface). In the case of small deformations, $n$ is approximately directed along $Ox$, $\frac{1}{R_1} + \frac{1}{R_2} \approx \partial_{yy} u_x$, and these equations lead to
\begin{align}
&\sigma_{xy} = 0, \label{equtangent}\\
&\hat{\sigma_{\rm s}}\, \partial_{yy} u_x + \sigma_{xx} + p = 0. \label{equnormal}
\end{align}
Although we consider the case of no fluid pressure on the surface (i.e., $p = 0$), we will represent the tension $\sigma_{\rm l}$ concentrated on the contact line as a Dirac distribution of ``pressure'' $p = -\sigma_{\rm l}\, \delta(y)$, which allows to include Eq.~(\ref{equline}) in Eq.~(\ref{equnormal}), with the unique equation on the surface
\begin{eqnarray}
\hat{\sigma_{\rm s}}\, \partial_{yy} u_x + \sigma_{xx} = \sigma_{\rm l}\, \delta(y). \label{equnormalgeneral}
\end{eqnarray}
Indeed, integrating this equation on the surface, for $-\varepsilon \leq y \leq \varepsilon$, and taking $\varepsilon  \rightarrow 0$, gives
\begin{eqnarray*}
2 \hat{\sigma_{\rm s}}\, \partial_y u_x(0,0+) + \lim_{\varepsilon \rightarrow 0} \int_{-\varepsilon}^{\varepsilon} \sigma_{xx} dy = \sigma_{\rm l}, 
\end{eqnarray*}
which leads to Eq.~(\ref{equline}), because the term involving $\sigma_{xx}$ is equal to 0, as explained in the preceding section: it is a consequence of Eq.~(\ref{inequality}) or Eq.~(\ref{linevolumestress}) (which will be confirmed for the solution of the present paper) and expresses that the volume stress $\sigma_{xx}$ gives no contribution at the contact line.

\section{The analytic functions $F$ and $H$} \label{FH}

In the following, $z$ will denote the complex variable $x + iy$ and $u$ the complex displacement $u_x + iu_y$ (function of the complex variable $z$). We use the general Kolosov's solution of plane strain elasticity
\begin{eqnarray*}
u(z) &= - \displaystyle\frac{1}{2\mu}\,(k\,F(z) + z\,\overline{F'(z)} + \overline{G(z)}),
\end{eqnarray*}
where $k = - \frac{\lambda + 3\mu}{\lambda + \mu} = 4\nu - 3$ ($\lambda$, $\mu$ Lam\'e's coefficients, $\nu$ Poisson's coefficient; $-3 < k < -1$; see Ref.~\onlinecite{Mandel:1966}, vol. II, ann. XVI), based on the two analytic functions $F$ and $G$, which we will write using the new analytic function $H = G - zF'$ (i.e., $H(z) = G(z) - zF'(z)$):
\begin{eqnarray}
-2\mu\, u = k\,F + \overline{H} + (z + \bar z)\,\overline{F'}.  \label{u}
\end{eqnarray}
As consequences,
\begin{align}
-2\mu\, \partial_x u &= k\,F' + \overline{H'} + 2\overline{F'} + (z + \bar z)\,\overline{F''} \nonumber\\
-2\mu\, \partial_y u &= i(k\,F' - \overline{H'} - (z + \bar z)\,\overline{F''}) \nonumber\\
-2\mu\, \partial_{yy} u &= -(k\,F'' + \overline{H''} + (z + \bar z)\,\overline{F'''}) \label{du}.
\end{align}
Similarly, Kolosov's expressions of the volume stresses may be written as
\begin{align}
\sigma_{xx} +i\sigma_{xy} &= F' - \overline{H'} - (z + \bar z)\,\overline{F''} \nonumber\\
\sigma_{yy} -i\sigma_{xy} &= F' + \overline{H'} + 2\overline{F'} + (z + \bar z)\,\overline{F''} \label{sigma}.
\end{align}

Owing to the symmetry $u(\bar z)= \overline{u(z)}$ of our problem, we look for analytic functions $F$ and $H$ with the same property, $F(\bar z)= \overline{F(z)}$ and $H(\bar z)= \overline{H(z)}$.

As we expect that $\sigma_{xx}$, $\sigma_{xy}$, and $\partial_{yy} u_x$ vanish at the infinity of the body, i.e., for $|z| \rightarrow +\infty$ with $x > 0$ (this will be confirmed on the final solution), Eqs.~(\ref{equtangent}) and (\ref{equnormalgeneral}) on the surface $x = 0$ may be extended at the infinity of the body: 
\begin{align}
&\sigma_{xy} = 0 \quad \text{on $x = 0$ and at the infinity}, \label{sigmaxy}\\
&\hat{\sigma_{\rm s}}\, \partial_{yy} u_x + \sigma_{xx} 
= \begin{cases}
\sigma_{\rm l}\, \delta(y) \quad \text{on $x=0$},\\
0 \quad \text{at the infinity}, \label{sigmaxx}
\end{cases}
\end{align}
i.e., using the above expressions
\begin{align*}
&\Im(F' - \overline{H'} - (z + \bar z)\,\overline{F''}) = 0 \;\text{on $x=0$ and}\,\rm{at\,the\,infinity}, \\
&\Re(F' - \overline{H'} - (z + \bar z)\,\overline{F''}) \\
&+ \frac{\hat{\sigma_{\rm s}}}{2\mu}\, \Re(k\,F'' + \overline{H''} + (z + \bar z)\,\overline{F'''}) \\
&= \begin{cases}
\sigma_{\rm l}\, \delta(y) \quad \text{on $x=0$},\\
0 \quad \text{at the infinity}.
\end{cases}
\end{align*}
Assuming that $(z + \bar z)\,F''$ and $\Re((z + \bar z)\,F''')$ vanish at the infinity (which will be confirmed on the final solution) and since $z + \bar z = 0$ on $x = 0$, the preceding equations may be written as
\begin{align}
&\Im(F' + H') = 0 \quad \text{on $x=0$ and at the infinity}, \label{equFH1}\\
&\Re(F' - H' + \frac{\hat{\sigma_{\rm s}}}{2\mu} (k\,F'' + H''))
= \begin{cases}
\sigma_{\rm l}\, \delta(y) \quad \text{on $x=0$},\\
0 \quad \text{at the infinity}. \label{equFH2}
\end{cases}
\end{align}
Using the transformation $z = \omega (\zeta) = \displaystyle \frac{1 - \zeta}{1 + \zeta}$ and denoting $\Phi(\zeta) = F(\omega (\zeta))$, $\Psi(\zeta) = H(\omega (\zeta))$, $\rm B$ the disk $|\zeta| < 1$, and $\rm S$ the circle $|\zeta| = 1$, Eq.~(\ref{equFH1}) becomes
\begin{eqnarray*}
\Im(-\frac{(1 + \zeta)^2}{2}(\Phi' + \Psi')) = 0 \quad \text{on $\rm S$},
\end{eqnarray*}
and, since the first member is an harmonic function in $\rm B$, it leads to
\begin{eqnarray*}
\Im(-\frac{(1 + \zeta)^2}{2}(\Phi' + \Psi')) = 0 \quad \text{in $\rm B$}.
\end{eqnarray*}
The analytic function $-\frac{(1 + \zeta)^2}{2}(\Phi' + \Psi')$ is therefore equal to a real constant $a$ in $\rm B$, hence
\begin{eqnarray}
F' + H' = \text{real const. $a$ (for $\Re z \geq 0$)}. \label{equH}
\end{eqnarray}
Eq.~(\ref{equFH2}) then becomes
\begin{eqnarray*}
\Re(F' + \frac{\hat{\sigma_{\rm s}}}{4\mu} (k - 1)F''-\frac{a}{2})
= \begin{cases}
\frac{1}{2}\,\sigma_{\rm l}\, \delta(y) \quad \text{on $x=0$},\\
0 \quad \text{at the infinity},
\end{cases}
\end{eqnarray*}
i.e., using the variable $\zeta$,
\begin{align*}
&\Re(\Xi(\zeta)) = \sigma_{\rm l}\, \delta(\theta) \quad \text{on $\rm S$, where}\\
&\Xi(\zeta) = -\frac{(1 + \zeta)^2}{2}\,\Phi' - \alpha \frac{(1 + \zeta)^3}{2}(\Phi' + \frac{1 + \zeta}{2}\,\Phi'') -\frac{a}{2},
\end{align*}
$\alpha = \displaystyle \frac{\hat{\sigma_{\rm s}}}{4\mu} (1 - k) > 0$, and $\zeta = e^{i\theta}$, $-\pi \leq \theta < \pi$ ($\delta(\theta) = \frac{1}{2}\,\delta(y)$ because, when $x = 0$, $d\theta = -2\, dy$ at $y = 0$). The function $\Re(\Xi(\zeta))$ being harmonic in $\rm B$, we apply Poisson's formula (see Ref.~\onlinecite{Dieudonne:1977}, (23.61.11.2))
\begin{align*}
\Re(\Xi(\zeta)) &= \frac{1}{2\pi} \int_{-\pi}^{\pi} \frac{1 - |\zeta|^2}{|\zeta - e^{i\theta}|^2}\, \sigma_{\rm l}\, \delta(\theta)\, d\theta \\ 
&= \frac{\sigma_{\rm l}}{2\pi}\, \frac{1 - |\zeta|^2}{|\zeta - 1|^2} \\
&= \frac{\sigma_{\rm l}}{2\pi}\, \Re \Big( \frac{1 + \zeta}{1 - \zeta} \Big) \quad \text{in $\rm B$},
\end{align*}
i.e.,
\begin{eqnarray*}
\Re \big( \Xi(\zeta) - \beta\, \displaystyle \frac{1 + \zeta}{1 - \zeta} \big) = 0  \quad \text{in $\rm B$},
\end{eqnarray*}
with $\beta = \displaystyle \frac{\sigma_{\rm l}}{2\pi}$. The analytic function $\Omega(\zeta) = \Xi(\zeta) - \beta\, \displaystyle \frac{1 + \zeta}{1 - \zeta}$ is therefore equal to a pure imaginary constant in $\rm B$, and thus equal to 0 (owing to the symmetry property of $\Omega$, $\Omega(\bar \zeta)= \overline{\Omega(\zeta)}$, due to the same property of $\Phi$):
\begin{eqnarray*}
\Xi(\zeta) - \beta\, \displaystyle \frac{1 + \zeta}{1 - \zeta} = 0  \quad \text{in $\rm B$},
\end{eqnarray*}
i.e., with the variable $z$,
\begin{eqnarray*}
F' - \alpha F''-\frac{a}{2} - \frac{\beta}{z} = 0 \quad \text{for $\Re z \geq 0$ (and $z \neq 0$)},
\end{eqnarray*}
which implies that $F - \alpha F'-\displaystyle \frac{a z}{2} - \beta \log z$ is a real constant (since $F(\bar z)= \overline{F(z)}$, implying the same property for the preceding expression; we use the usual logarithm defined in $\bf C - R_-$). We may take this constant equal to $0$, since an additive constant in $F$ (or in $H$) only produces an additive constant in $u$ but does'nt change the derivatives of $u$ and the stress tensor (according to Eqs.~(\ref{u})--(\ref{sigma})):
\begin{align}
F - \alpha F'-\frac{a z}{2} - \beta \log z = 0 \quad \text{for $\Re z \geq 0$ ($z \neq 0$)}. \label{equF}
\end{align}

\section{The solution} \label{SOL}

By extension of the known solution of the differential equation~(\ref{equF}) when $z$ is a real variable (see Ref.~\onlinecite{Bourbaki:1976}, chap. IV, \S § 2, $\rm n^{\circ}$ 3), we obtain the general solution of this equation when $z$ is a complex variable, as
\begin{eqnarray*}
F(z) = b\, e^{z/\alpha} - e^{z/\alpha} P(z),
\end{eqnarray*}
where $P(z)$ is a primitive in $\bf C - R_-$ of the function $e^{-z/\alpha} (\frac{a}{2\alpha} z + \frac{\beta}{\alpha} \log z)$, satisfying $P(\bar z)= \overline{P(z)}$, and $b$ a real constant (since $F(\bar z)= \overline{F(z)}$). We then find
\begin{eqnarray*}
P(z) = -\frac{a}{2}\, e^{-z/\alpha} (z + \alpha) - \beta (e^{-z/\alpha} \log z - {\rm Ei}(-\frac{z}{\alpha})),
\end{eqnarray*}
where ${\rm Ei}$ is the ``exponential integral'' function,\cite{Wikipedia} which we here define as the primitive of the function $\displaystyle \frac{e^z}{z}$ in $\bf C - R_+$, which coincides with the function $x \rightarrow \int_{-\infty}^x \frac{e^t}{t} dt$ when $x \in \bf R_-^*$. This gives (after addition of the constant $-\beta \log \alpha$)
\begin{eqnarray*}
F(z) = b\, e^{z/\alpha} + \frac{a}{2} (z + \alpha) + \beta (\log \frac{z}{\alpha} - e^{z/\alpha}\, {\rm Ei}(-\frac{z}{\alpha})).
\end{eqnarray*}
From this expression, we obtain
\begin{align}
(z + \bar z)\,F'' =\, &\frac{b}{\alpha^2} (z + \bar z)\, e^{z/\alpha} \nonumber\\
&- \frac{\beta}{\alpha}\, \frac{z + \bar z}{z} (\frac{z}{\alpha}\, e^{z/\alpha}\, {\rm Ei}(-\frac{z}{\alpha}) + 1),\label{zF''}
\end{align}
which shows that $b$ must be equal to 0, in order to satisfy the first assumption preceding Eqs.~(\ref{equFH1})-(\ref{equFH2}), because $\frac{z}{\alpha}\, e^{z/\alpha}\, {\rm Ei}(-\frac{z}{\alpha}) + 1 \rightarrow 0$ at the infinity, for $\Re z > 0$ (proved in the Appendix). Note that (with $b = 0$)
\begin{eqnarray*}
(z + \bar z)\,F''' = -\frac{\beta}{\alpha^2} \frac{z + \bar z}{z} (\frac{z}{\alpha}\, e^{z/\alpha}\, {\rm Ei}(-\frac{z}{\alpha}) + 1)
+ \frac{\beta}{\alpha} \frac{z + \bar z}{z^2}
\end{eqnarray*}
also tends to 0 at the infinity, satisfying the second assumption.

Moreover, according to Eqs.~(\ref{sigma}), (\ref{du}) and (\ref{equH}),
\begin{align*}
\sigma_{xx} +i\sigma_{xy} &= \Re (2F' - a) - (z + \bar z)\,\overline{F''}\\ 
&= \Re (-\frac{2\beta}{\alpha }\, e^{z/\alpha}\, {\rm Ei}(-\frac{z}{\alpha})) - (z + \bar z)\,\overline{F''},\\
\sigma_{yy} -i\sigma_{xy} &= \Re (2F' - a) + 2a + (z + \bar z)\,\overline{F''},\\
-2\mu\, \partial_{yy} u &= -(k\,F'' - \overline{F''} + (z + \bar z)\,\overline{F'''}),
\end{align*}
which shows that $\sigma_{xx}$, $\sigma_{xy}$, and $\partial_{yy} u$ tend to 0 and $\sigma_{yy}$ tends to $2a$ at the infinity (because $e^{z/\alpha}\, {\rm Ei}(-\frac{z}{\alpha}) \sim -\frac{\alpha}{z}$: see Appendix). The assumptions preceding Eqs.~(\ref{sigmaxy})-(\ref{sigmaxx}) are thus satisfied. In the following, we take $a = 0$, in order that all the components of the stress tensor vanish at the infinity (including $\sigma_{z'z'} = \nu (\sigma_{xx} + \sigma_{yy})$). According to $H = -F$ (from Eq.~(\ref{equH}), taking the integration constant equal to 0), the final solution is then
\begin{align}
-2\mu\, u &= k\,F - \overline{F} + (z + \bar z)\,\overline{F'}, \nonumber\\
\text{where }F(z) &= \beta (\log \frac{z}{\alpha} - e^{z/\alpha}\, {\rm Ei}(-\frac{z}{\alpha})).\label{Fsol}
\end{align}

The preceding solution holds for $\alpha > 0$. Note that, if $\alpha = 0$ (i.e., $\hat{\sigma_{\rm s}} = 0$), we obtain the solution
\begin{align}
-2\mu\, u &= k\,F - \overline{F} + (z + \bar z)\,\overline{F'}, \nonumber\\
\text{with }F(z) &= \beta \log z, \label{Flamant}
\end{align}
from Eq.~(\ref{equF}) (with $a = 0$, for the same reason as above), which is exactly the classical Flamant's solution.\cite{Flamant:1892}

\section{Displacements and stresses}

From Eqs.~(\ref{du})-(\ref{sigma}) (with $H = -F$) and the above expression of $F$ (Eq.~(\ref{Fsol})), we also have
\begin{align}
-2\mu\, \partial_x u &= k\,F' + \overline{F'} + (z + \bar z)\,\overline{F''} \nonumber\\
-2\mu\, \partial_y u &= i(k\,F' + \overline{F'} - (z + \bar z)\,\overline{F''}) \nonumber\\
\sigma_{xx} +i\sigma_{xy} &= F' + \overline{F'} - (z + \bar z)\,\overline{F''} \nonumber\\
\sigma_{yy} -i\sigma_{xy} &= F' + \overline{F'} + (z + \bar z)\,\overline{F''}, \quad \text{with} \nonumber\\
F'(z) &= -\frac{\beta}{\alpha}\, e^{z/\alpha}\, {\rm Ei}(-\frac{z}{\alpha}) \nonumber\\
F''(z) &= -\frac{\beta}{\alpha^2} (e^{z/\alpha}\, {\rm Ei}(-\frac{z}{\alpha}) + \frac{\alpha}{z}) \label{dusigmadFsol}.
\end{align}

Introducing the function $E(z) = \log z - e^z {\rm Ei}(-z)$, we may write
\begin{eqnarray}
u(z) = \frac{\beta}{2\mu} (-k\,E(\frac{z}{\alpha}) + \overline{E}(\frac{z}{\alpha}) - \frac{z + \bar z}{\alpha}\, \overline{E'}(\frac{z}{\alpha})),\label{u(z/alpha)}
\end{eqnarray}
which shows that the displacement $u$ is proportional to $\beta/(2\mu)$ and $(2\mu/\beta) u$ only depends on $z/\alpha$ and $k$ (i.e., $\nu$). Thus, $\alpha = \hat{\sigma_{\rm s}}(1 - \nu)/\mu$ and $\beta/(2\mu) = \sigma_{\rm l}/(4\pi\mu)$ are the characteristic elastocapillary lengths for this solution. Similarly, $(2\mu\alpha/\beta) \partial_x u$ and $(2\mu\alpha/\beta) \partial_y u$ only depend on $z/\alpha$ and $k$, and $(\alpha/\beta) \sigma_{xx}$, $(\alpha/\beta) \sigma_{yy}$, and $(\alpha/\beta) \sigma_{xy}$ only depend on $z/\alpha$.

Since $\frac{e^z}{z} = \frac{1}{z} + \overset{\infty}{\underset{n = 1}{\sum}} \frac{z^{n-1}}{n!}$, we have ${\rm Ei}(z) = 
\log(-z) + \overset{\infty}{\underset{n = 1}{\sum}} \frac{z^n}{n\, n!} + \text{const.}\,$ in $\bf C - R_+$, the constant being equal to the Euler constant $\gamma$ because ${\rm Ei}(x) = \int_{-\infty}^x \frac{e^t}{t} dt$ when $x \in \bf R_-^*$:
\begin{eqnarray*}
{\rm Ei}(z) = \gamma + \log(-z) + \overset{\infty}{\underset{n = 1}{\sum}}\, \frac{z^n}{n\, n!}  \quad \text{in $\bf C - R_+$}.
\end{eqnarray*}
Then, in $\bf C - R_-$,
\begin{eqnarray*}
E(z) = -\gamma\, e^z + (1 - e^z)\log z - e^z\, \overset{\infty}{\underset{n = 1}{\sum}}\, \frac{(-z)^n}{n\, n!}
\end{eqnarray*}
tends to $-\gamma$ and
\begin{align*}
(z + \bar z) E'(z) = &-e^z (z + \bar z)\log z \\
&-e^z (z + \bar z)(\gamma + \overset{\infty}{\underset{n = 1}{\sum}}\, \frac{(-z)^n}{n\, n!})
\end{align*}
tends to 0, when $z \rightarrow 0$, which leads to the finite limit for the displacement
\begin{eqnarray}
\lim_{z \rightarrow 0} u(z) = -\frac{\beta}{2\mu} (1 - k)\gamma = -\frac{\gamma}{\pi} \frac{\sigma_{\rm l}}{\mu} (1 - \nu).\label{u(0)}
\end{eqnarray}

\begin{figure}[htbp]
\begin{center}
\includegraphics[width=16cm]{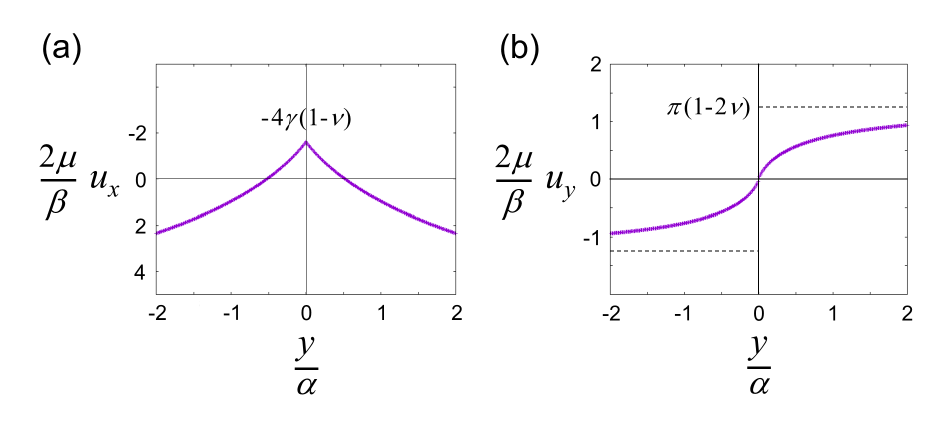}
\end{center}
\caption{Displacements $u_x$ (a) and $u_y$ (b) on the surface $x = 0$ as functions of $y$, for $\nu = 0.3$. The graph $u_x$ (a) represents the form of the surface after deformation. Notations: $\alpha = \displaystyle\frac{\hat{\sigma_{\rm s}}(1 - \nu)}{\mu} $, $\beta = \displaystyle\frac{\sigma_{\rm l}}{2\pi}$.} \label{usurf}
\end{figure}

\begin{figure}[htbp]
\begin{center}
\includegraphics[width=16cm]{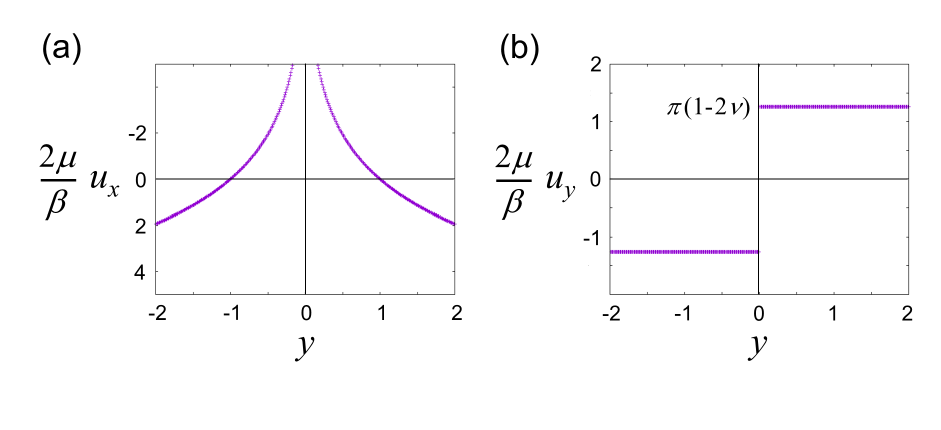}
\end{center}
\caption{The classical Flamant's solution: displacements $u_x$~(a) and $u_y$ (b) on the surface $x = 0$ as functions of $y$, for $\nu = 0.3$. Notation: $\beta = \displaystyle\frac{\sigma_{\rm l}}{2\pi}$. The graph $u_x$ (a) represents the form of the surface after deformation. At $x = y = 0$, $u_x$ is infinite and $u_y$ is discontinuous.} \label{usurfFlamant}
\end{figure}

From the above expressions of $F'(z)$, $F''(z)$, and ${\rm Ei}(z)$, we obtain
\begin{align*}
\frac{\alpha}{\beta}\, F'(z) &= -\gamma - \log \frac{z}{\alpha} + \varepsilon_0(z) \\
&= -\gamma - \log \frac{r}{\alpha} - i\theta + \varepsilon_0(z),\\
-\frac{\alpha}{\beta} (z + \bar z) F''(z) &= \frac{z + \bar z}{z} + \varepsilon_0(z) = 1 + e^{-2i\theta} + \varepsilon_0(z),
\end{align*}
where $z = r e^{i\theta}$, $r > 0$, $-\pi < \theta < \pi$, and $\varepsilon_0(z)$ denotes any expression of $z$ which tends to 0 when $z \rightarrow 0$. Thus,
\begin{align}
\frac{2\mu\alpha}{\beta}\, \partial_y u = \,&(1 - k)\theta + \sin 2\theta \nonumber\\ 
&+ i((1 + k)(\gamma + \log \frac{r}{\alpha}) - 1 - \cos 2\theta) + \varepsilon_0(z),\label{dyu(vois0)}
\end{align}
which shows that
\begin{align}
\lim_{z \rightarrow 0,\; \theta\;\text{constant}} \partial_y u_x &= \frac{\beta}{2\mu\alpha}((1 - k)\theta + \sin 2\theta) \nonumber\\
&= \frac{\sigma_{\rm l}}{\hat{\sigma_{\rm s}}} (\frac{\theta}{\pi} + \frac{\sin 2\theta}{4\pi(1 - \nu)}),\nonumber\\
\lim_{z \rightarrow 0} \partial_y u_y &=
\begin{cases}
+\infty \quad \text{if $\sigma_{\rm l} > 0$}\\
-\infty \quad \text{if $\sigma_{\rm l} < 0$}.\label{dyu(0)}
\end{cases}
\end{align}
Obviously, the infinite limit of $\partial_y u_y$ (and, below, $\partial_x u_x$) is in contradiction with the assumption of small deformations and, strictly speaking, the solution is not valid for $z$ close to 0. The solution may rather be considered as a mathematical description of the singularity at $z = 0$. Note that, on the surface $x = 0$, with $y > 0$ (i.e., $\theta = \pi/2$), the above equation shows that $\partial_y u_x$ tends to 
$\displaystyle\frac{\sigma_{\rm l}}{2\hat{\sigma_{\rm s}}}$ when $y \rightarrow 0$, $y > 0$, in agreement with the above Eq.~(\ref{equline}). Since the initial surface $x = 0$ is, after deformation, represented by the function $y \rightarrow u_x(0,y)$, there is thus a finite displacement and the formation of a ridge at $z = 0$ on the surface, as shown in Fig.~\ref{usurf}a.
In this Fig.~\ref{usurf}, the displacements $u_x(0,y)$ and $u_y(0,y)$ on the surface, obtained from Eq.~(\ref{Fsol}), are represented, for $\nu = 0.3$. For comparison, the displacements of the Flamant's solution (Eq.~(\ref{Flamant})) are shown in Fig.~\ref{usurfFlamant}.
Note that, except for the presence of a ridge at the contact line, the behavior on the surface of our solution (Fig.~\ref{usurf}) cannot be compared with the results of Ref.~\onlinecite{Jerison-etal:2011} which concern a thin film $0 \leq x \leq h$ of elastic body with a zero-displacement at $x = h$.

In a similar way, we have
\begin{align}
\frac{2\mu\alpha}{\beta}\,\partial_x u = \,&(1 + k)(\gamma + \log \frac{r}{\alpha}) + 1 + \cos 2\theta \nonumber\\ 
&+ i(-(1 - k)\theta + \sin 2\theta) + \varepsilon_0(z),\label{dxu(vois0)}
\end{align}
which shows that
\begin{align}
\lim_{z \rightarrow 0} \partial_x u_x &=
\begin{cases}
+\infty \quad \text{if $\sigma_{\rm l} > 0$}\\
-\infty \quad \text{if $\sigma_{\rm l} < 0$},
\end{cases} \nonumber\\
\lim_{z \rightarrow 0,\; \theta\;\text{constant}} \partial_x u_y 
&= \frac{\sigma_{\rm l}}{\hat{\sigma_{\rm s}}} (-\frac{\theta}{\pi} + \frac{\sin 2\theta}{4\pi(1 - \nu)}),\label{dxu(0)}
\end{align}
and immediately gives the limits of the strain tensor components $\varepsilon_{xx} = \partial_x u_x$, $\varepsilon_{yy} = \partial_y u_y$, and $\varepsilon_{xy} = \frac{1}{2}(\partial_x u_y + \partial_y u_x)$: 
\begin{eqnarray}
\lim_{z \rightarrow 0,\; \theta\;\text{constant}} \varepsilon_{xy} = \frac{\sigma_{\rm l}}{\hat{\sigma_{\rm s}}}\, \frac{\sin 2\theta}{4\pi(1 - \nu)}.\label{epsilon(0)}
\end{eqnarray}

We also obtain
\begin{align}
\frac{\alpha}{\beta}(\sigma_{xx} +i\sigma_{xy}) = \,&-2(\gamma + \log \frac{r}{\alpha}) + 1 + \cos 2\theta \nonumber\\ 
&+ i \sin 2\theta + \varepsilon_0(z), \nonumber\\
\frac{\alpha}{\beta}(\sigma_{yy} -i\sigma_{xy}) = \,&-2(\gamma + \log \frac{r}{\alpha}) - 1 - \cos 2\theta \nonumber\\
&- i \sin 2\theta + \varepsilon_0(z), \label{sigma(vois0)}
\end{align}
which shows that
\begin{align}
\lim_{z \rightarrow 0} \sigma_{xx} &=
\begin{cases}
+\infty \quad \text{if $\sigma_{\rm l} > 0$}\\
-\infty \quad \text{if $\sigma_{\rm l} < 0$},
\end{cases} \nonumber\\
\lim_{z \rightarrow 0} \sigma_{yy} &=
\begin{cases}
+\infty \quad \text{if $\sigma_{\rm l} > 0$}\\
-\infty \quad \text{if $\sigma_{\rm l} < 0$},
\end{cases} \nonumber\\
\lim_{z \rightarrow 0,\; \theta\;\text{constant}} \sigma_{xy} 
&= \frac{\sigma_{\rm l}}{\hat{\sigma_{\rm s}}}\, \frac{\mu}{2\pi(1 - \nu)}\, \sin 2\theta. \label{sigma(0)}
\end{align}

Since the inequality $|1 + z\, e^z {\rm Ei}(-z)| \leq \frac{2}{|z|}$ for $\Re z > 0$ (proved in the Appendix) may be continuously extended in $\Re z \geq 0$, $z \neq 0$ (let us recall that $z \rightarrow {\rm Ei}(-z)$ is analytic in $\bf C - R_-$), then $e^z {\rm Ei}(-z) \sim -\frac{1}{z}$ when \mbox{$|z| \rightarrow +\infty$}, \mbox{$\Re z \geq 0$}, and we may write (from Eqs.~(\ref{Fsol}) and (\ref{dusigmadFsol}))
\begin{align*}
\frac{1}{\beta}\, F(z) &= \log \frac{r}{\alpha} + i\theta + \varepsilon_\infty(z),\\
\frac{1}{\beta} (z + \bar z) F'(z) &= -\frac{z + \bar z}{z}\, \frac{z}{\alpha}\, e^{z/\alpha}\, {\rm Ei}(-\frac{z}{\alpha})\\ 
&= 1 + e^{-2i\theta} + \varepsilon_\infty(z),
\end{align*}
where $\varepsilon_\infty(z)$ denotes any expression of $z$ which tends to 0 when $|z| \rightarrow +\infty$, $\Re z \geq 0$. Thus,
\begin{align*}
\frac{2\mu}{\beta}\, u = \,&(1 - k)\log \frac{r}{\alpha} - 1 - \cos 2\theta\\
&+ i(-(1 + k)\theta - \sin 2\theta) + \varepsilon_\infty(z),
\end{align*}
which shows that
\begin{align}
\lim_{|z| \rightarrow +\infty} u_x &=
\begin{cases}
+\infty \quad \text{if $\sigma_{\rm l} > 0$}\\
-\infty \quad \text{if $\sigma_{\rm l} < 0$},
\end{cases} \nonumber\\
\lim_{|z| \rightarrow +\infty,\; \theta\;\text{constant}} u_y 
&= \frac{\sigma_{\rm l}}{4\pi\mu} (2(1 - 2\nu)\theta - \sin 2\theta).\label{u(infini)}
\end{align}
Clearly, the infinite limit of $u_x$ is due to the assumption of a force $\sigma_{\rm l}$ applied on the whole infinite line $x = y = 0$ (the $z'$ axis). Note that, on the surface $x = 0$, with $y > 0$ (i.e., $\theta = \pi/2$), the preceding equation shows that $u_y$ tends to $\displaystyle\frac{\sigma_{\rm l}}{4\mu}(1 - 2\nu)$ when $y \rightarrow +\infty$ (see Fig.~\ref{usurf}b). Also note that, in the limit case $\nu \rightarrow \frac{1}{2}$ (i.e., $k + 1 \rightarrow 0$), the expression of $u$ (Eq.~(\ref{Fsol})) shows that $u_y = 0$ on the surface $x = 0$.

From the expression of $F'(z)$ in Eq.~(\ref{dusigmadFsol}), $F'(z)$ tends to 0 and, from Eq.~(\ref{zF''}) (with $b = 0$), $(z + \bar z)\,F''(z)$ tends to 0 when $|z| \rightarrow +\infty$, $\Re z \geq 0$, which implies that
\begin{align}
\partial_x u \;\,\text{and }\,\partial_y u \;\,\text{tend to 0 when }|z| \rightarrow +\infty, \label{du(infty)}
\end{align}
and
\begin{align}
\sigma_{xx}, \sigma_{xy}, \sigma_{yy}, \text{and }\sigma_{z'z'}\;\text{tend to 0 when }|z| \rightarrow +\infty \label{sigma(infty)}
\end{align}
(as mentioned in Sec.~\ref{SOL}).

The dependence on $\nu$ is shown in Eq.~(\ref{u(z/alpha)}): at $z/\alpha$ fixed, $(2\mu/\beta) u$ linearly varies with $k$, i.e., with $\nu$. In particular, on the surface $x = 0$,
\begin{align*}
\frac{2\mu}{\beta}\, u_x(0,y) &= 4(1 - \nu)\, \Re(E(\frac{iy}{\alpha})) \\
\frac{2\mu}{\beta}\, u_y(0,y) &= 2(1 - 2\nu)\, \Im(E(\frac{iy}{\alpha})),
\end{align*}
which shows that all the functions $y/\alpha \rightarrow (2\mu/\beta) u_x(0,y)$ and $y/\alpha \rightarrow (2\mu/\beta) u_y(0,y)$ are similar to those of Fig.~\ref{usurf}, i.e., only multiplied by $\frac{1 - \nu}{1 - 0.3}$ for $u_x$, and by $\frac{1 - 2\nu}{1 - 0.6}$ for $u_y$.

\begin{figure}[htbp]
\begin{center}
\includegraphics[width=16cm]{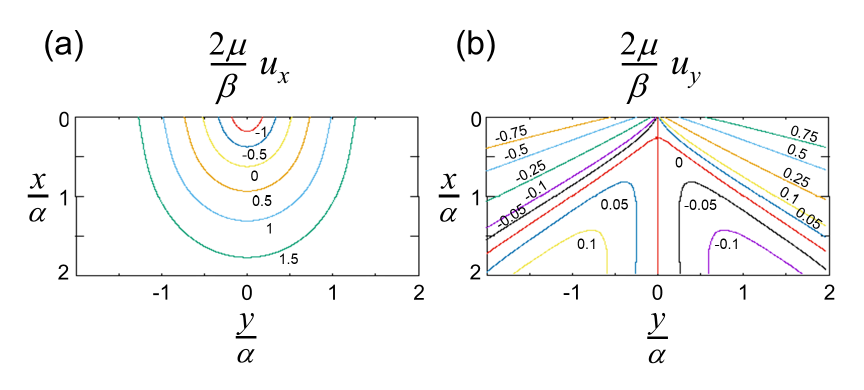}
\end{center}
\caption{Displacements $u_x$ (a) and $u_y$ (b) in the volume $x \geq 0$ as functions of $(x,y)$, for $\nu = 0.25$. Vertical axis $x/\alpha$, horizontal axis $y/\alpha$. The values of  $(2\mu/\beta) u_x$ or $(2\mu/\beta) u_y$ are indicated on each curve. Notations: $\alpha = \displaystyle\frac{\hat{\sigma_{\rm s}}(1 - \nu)}{\mu} $, $\beta = \displaystyle\frac{\sigma_{\rm l}}{2\pi}$.} \label{uvol}
\end{figure}

In the volume $x \geq 0$, the displacements $u_x$ and $u_y$ are represented in Fig.~\ref{uvol}, for $\nu = 0.25$.
In Fig.~\ref{uvol}a, we observe that $u_x$ increases when the distance to $(x,y) = (0,0)$ increases. In the $xy$ plane of Fig.~\ref{uvol}b, the curves \mbox{$u_y$ = constant} show asymptotic directions. There are two lines where $u_y = 0$: the straight line $y = 0$ (obviously) and a curved line, with an intersection of the two lines at $x_0 > 0$, $y = 0$. If $x$ remains constant, $x > x_0$, and $y$ increases from 0 to positive values, we observe that $u_y$ decreases from 0 to a negative value and then increases to positive values (after crossing the value 0). These observations are consistent with Eq.~(\ref{u(infini)}) which gives the asymptotic value $u_{y,\infty}$ of $u_y$ as a function of the direction $\theta$: $(2\mu/\beta) u_{y,\infty} =$ \mbox{$2(1 - 2\nu)\theta - \sin 2\theta$}, which decreases from 0 to a negative value and then increases to the positive value $\pi(1 - 2\nu)$ (after crossing the value 0), when $\theta$ increases from 0 to $\pi/2$. In fact, the asymptotic direction(s) $\theta$ of the curve $(2\mu/\beta) u_y = c$ (constant) is (are) given by the equation $c = 2(1 - 2\nu)\theta - \sin 2\theta$. Thus, for $y \geq 0$, the curves $(2\mu/\beta) u_y = c$ with $c > 0$ have one asymptotic direction and those with $c \leq 0$ have two asymptotic directions, which is consistent with Fig.~\ref{uvol}b. Note that these results concerning $u_y$ are consistent with the experiments and Fourier transform calculations of Ref.~\onlinecite{Kim-etal:2021}-Fig.~3a--d, in the neighborhood of the contact line.

As $\nu$ increases, the distances between the curves of Fig.~\ref{uvol}a increase, which means that (at $2\mu\alpha/\beta$ constant) the gradient of $u_x$ decreases, at each point $(x/\alpha,y/\alpha)$. In Fig.~\ref{uvol}b, as $\nu$ increases to 0.5, $x_0/\alpha$ (abscissa of the intersection of the straight line $y = 0$ and the curved line $u_y = 0$) decreases to 0. Moreover, if $\theta_0$ denotes the asymptotic direction of the curved line $u_y = 0$, i.e., $2(1 - 2\nu)\theta_0 = \sin 2\theta_0$, $0 \leq \theta_0 \leq \pi/2$, clearly $\theta_0$ increases from 0 to $\pi/2$ when $\nu$ increases from 0 to 0.5. When $\nu$ reaches the limit value 0.5, this curved line $u_y = 0$ becomes the straight line $x = 0$.

\section{The displacements belong to $H^1(\rm V)$ and $H^1(\rm S)$: finite elastic energy and validity of Green's formulae}

Eqs.~(\ref{dyu(vois0)}), (\ref{dxu(vois0)}), and (\ref{sigma(vois0)}) show that $\partial_y u_x$, $\partial_x u_y$, and $\sigma_{xy}$ are bounded for $|z| < r_0$ ($r_0$ positive constant) and
\begin{align}
&|\partial_x u_x|, |\partial_y u_y|, |\sigma_{xx}|, |\sigma_{yy}|,\,\text{and }|\sigma_{z'z'}| \leq c|\log r| + d \nonumber\\
&\text{for }|z| < r_0 \label{inequal}
\end{align}
($c$ and $d$ positive constants), as in our preceding paper Ref.~\onlinecite{Olives:2014}. If $\rm V$ denotes the volume $x > 0$, $|x + iy| < r_0$, \mbox{$0 < z' < l_0$} ($l_0$ positive constant), the preceding inequalities imply that all the components $\partial_j u_i$ and $\sigma_{ij}$ belong to $L^2(\rm V)$ (because $r(\log r)^2$ and $r |\log r|$ are integrable on $[0, r_0]$). Since $u$ is continuous (with a finite limit at $z = 0$), this shows that {\it the components $u_i$ of the displacement belong to the Sobolev space $H^1(\rm V)$} (as a consequence, the components $\sigma_{ij} \in L^2(\rm V)$). This point is crucial for the physical validity of the solution because it implies that:

1) The elastic energy
\begin{align*}
\mu \int_{\rm V}\,(\underset{(i,j)}{\sum} \varepsilon_{ij}^2 + \frac{\nu}{1 - 2\nu} (\underset{i}{\sum} \varepsilon_{ii})^2)\,dv
\end{align*}
is finite.

2) Green's formula
\begin{align}
\int_{\rm V} \sigma : {\rm D}w \,dv = -\int_{\rm V} {\rm div} \sigma \cdot w \,dv - \int_{\rm S} (\sigma \cdot n) \cdot w \,da \label{Greenv}
\end{align}
is valid, because the components $w_i$ (as $u_i$) belong to $H^1(\rm V)$, the components $\sigma_{ij}$ belong to $L^2(\rm V)$, and the components $({\rm div} \sigma)_i$ belong to $L^2(\rm V)$ (owing to the equilibrium equation ${\rm div} \sigma = 0$; note that the derivatives $\partial_l \sigma_{ij}$ generally do not belong to $L^2(\rm V)$ and the components $\sigma_{ij}$ do not belong to $H^1(\rm V)$) (see Theorem 4.4.7 in Ref.~\onlinecite{Allaire:2012}).

Let us recall that, in the classical Flamant's solution, $u$ is infinite at the contact line, the derivatives $\partial_j u_i$ (and $\sigma_{ij}$) do not belong to $L^2(\rm V)$, and the elastic energy is infinite.

In the theory presented in Ref.~\onlinecite{Olives:2010a}, we also used Green's formula on the surface
\begin{align}
\int_{\rm S} \bar{\sigma_{\rm s}} : {\rm D_s}w \,da = -\int_{\rm S} {\rm div} \bar{\sigma_{\rm s}} \cdot w \,da - \int_{\rm C} (\bar{\sigma_{\rm s}} \cdot \nu) \cdot w \,dl \label{Greens}
\end{align}
(with the notations of Sec.~\ref{GEN} and $({\rm D_s}w)_{\beta i} = \partial_\beta w_i$), where $\rm S$ is a portion of the $\rm bf$ (or $\rm bf'$) surface adjacent to the contact line, $\rm C$ the boundary curve, $\nu$ the unit vector normal to this curve, tangent to the surface, and directed towards the interior of $\rm S$, $da$ an element of area, and $dl$ an element of length. The derivatives $\partial_\alpha u_i$, being bounded or subjected to the inequality of Eq.~(\ref{inequal}), belong to $L^2(\rm S)$ (because $(\log r)^2$ and $|\log r|$ are integrable on $[0, r_0]$), so that {\it the components $u_i$ belong to $H^1(\rm S)$}. Here also, applying the same Theorem as above, Green's formula is valid, because the components $w_i$ (as $u_i$) belong to $H^1(\rm S)$, the components $\bar{\sigma_{\rm s}}^{\beta i}$ belong to $L^2(\rm S)$ (since $\hat{\sigma_{\rm s}}$ is constant on the surface), and the components $({\rm div} \bar{\sigma_{\rm s}})^i$ belong to $L^2(\rm S)$ (the tangential component is equal to 0 and the normal component is $\hat{\sigma_{\rm s}}\, \partial_{yy} u_x = -\sigma_{xx}$, for $x = 0$ and $y \neq 0$, which satisfies the inequality of Eq.~(\ref{inequal}); see Sec.~\ref{Application} and Eq.~(\ref{equnormalgeneral})).

\section{Conclusions}

When an elastic body is in contact with two immiscible fluids (e.g., an elastic gel in contact with a liquid and the air), it is subjected to the fluid--fluid surface tension acting on the body--fluid--fluid triple contact line. This force concentrated on the contact line produces a singularity on the elastic body. In the simple case of a semi-infinite body, bounded by a plane, and a force concentrated on a straight line of this plane and normal to the plane, the classical solution of Flamant\cite{Flamant:1892} leads to an infinite displacement at this contact line. In the present paper, a new solution with a finite displacement at the contact line is obtained, by introducing a surface tension (i.e., isotropic surface stress) at the surface of the semi-infinite body and applying the general equilibrium equations (on the surface and at the contact line) of the previous work Ref.~\onlinecite{Olives:2010a}. Using Kolosov's approach of plane strain elasticity, in the complex plane, with a new function $H$ (Sec.~\ref{FH}), and the theory of analytic functions, we are led to a differential equation (Eq.~(\ref{equF})), the solution of which gives the explicit expression of the displacement $u$ (Eq.~(\ref{Fsol})), and then the expressions of the derivatives of the displacement and the strain and stress components (Eq.~(\ref{dusigmadFsol})). The displacement $u$ is proportional to $\beta/(2\mu) = \sigma_{\rm l}/(4\pi\mu)$ ($\sigma_{\rm l}$ the force or fluid--fluid surface tension concentrated on the contact line, $\mu$ the Lam\'e elastic shear modulus) and $(2\mu/\beta)u$ only depends on the coordinates $(x/\alpha, y/\alpha)$ and $\nu$ ($\alpha = \hat{\sigma_{\rm s}}(1 - \nu)/\mu$, $\hat{\sigma_{\rm s}}$ the surface tension at the surface of the body, $\nu$ the Poisson's coefficient). Thus, $\alpha$ and $\beta/(2\mu)$ are the characteristic elastocapillary lengths for this solution. The displacement is finite and continuous at the contact line, with the formation of a ridge (see Eq.~(\ref{u(0)}) and Fig.~\ref{usurf}a). At the contact line, the $xx$ and $yy$ strain and stress components become infinite, but their $xy$ components have different finite limits when approaching the contact line under different directions (Eqs.~(\ref{dyu(0)}), (\ref{dxu(0)}), (\ref{epsilon(0)}), and (\ref{sigma(0)})). Moreover, all the stress components and the derivatives of the displacement vanish at the infinity of the body (Eqs.~(\ref{du(infty)})-(\ref{sigma(infty)})). The components $u_x$ and $u_y$ of the displacement, as functions of the coordinates $(x, y)$, are shown in Fig.~\ref{uvol}.

Moreover, in the neighborhood of the contact line, the derivatives of the displacement are either bounded or present a logarithmic divergence (Eq.~(\ref{inequal}); as in the example of solution of Ref.~\onlinecite{Olives:2014}), which implies that the components $u_i$ of the displacement belong to $H^1(\rm V)$ and $H^1(\rm S)$ ($\rm V$ being a volume of the body and $\rm S$ a portion of surface of the body, in the neighborhood of the contact line). This is a crucial point for the physical validity of the solution because it implies that the elastic energy is finite and both Green's formulae in the volume and on the surfaces (Eqs.~(\ref{Greenv}) and (\ref{Greens}), which play a central role in the theory of Ref.~\onlinecite{Olives:2010a}) are valid. Note that, in the classical Flamant's solution, the components $u_i$ do not belong to $H^1(\rm V)$ and the elastic energy is infinite.

\appendix

\section{Expression of ${\rm Ei}(z)$ when $\Re z < 0$} \label{Ei}

Since $|e^{zt}| = e^{xt}$, the function $t \rightarrow \frac{e^{zt}}{t}$ is integrable in $[1, +\infty[$, for $x = \Re z < 0$, and we note $f(z) = -\int_1^{+\infty} \frac{e^{zt}}{t} dt$.

If $\Re z = x < x_0 < 0$, then $|\frac{e^{zt}}{t}| = \frac{e^{xt}}{t} < \frac{e^{x_0 t}}{t}$ which is integrable for $t \in [1, +\infty[$, which implies that $f$ is analytic in $\Re z < x_0$ (see Ref.~\onlinecite{Dieudonne:1982}, (13.8.6)(iii)) for all $x_0 < 0$, hence $f$ is analytic in $\Re z < 0$.

Moreover, for $z = x \in \bf R_-^*$, $f(x) = -\int_1^{+\infty} \frac{e^{xt}}{t} dt = \int_{-\infty}^x \frac{e^s}{s} ds = {\rm Ei}(x)$. The two analytic functions $f$ and $\rm Ei$ are then equal in $\Re z < 0$, i.e.,
\begin{align*}
{\rm Ei}(z) = -\int_1^{+\infty} \frac{e^{zt}}{t} dt = -\int_{\Gamma_z} \frac{e^v}{v} dv\quad \text{for $\Re z < 0$},
\end{align*}
where $\Gamma_z$ is the path: $t \in [1, +\infty[\, \rightarrow v = zt \in \bf C$.

\section{Behavior of $\rm Ei$ at the infinity, when $\Re z < 0$} \label{Ei-infinity}

Since $|\frac{e^{zt}}{zt}| = \frac{e^{xt}}{|z|t} \rightarrow 0$ when $t \rightarrow +\infty$, for $x = \Re z < 0$, the integration of $(\frac{e^v}{v})' = \frac{e^v}{v} - \frac{e^v}{v^2}$ along the path $\Gamma_z$ gives
\begin{align*}
0 - \frac{e^z}{z} = -{\rm Ei}(z) - \int_{\Gamma_z} \frac{e^v}{v^2}\, dv, \; \text{i.e.,}
\end{align*}
\begin{align*}
\frac{e^z}{z} -{\rm Ei}(z) &= \int_{\Gamma_z} \frac{e^v}{v^2}\, dv = \frac{1}{z} \int_1^{+\infty} \frac{e^{zt}}{t^2} dt = \frac{e^z}{z} I(z),\\
\text{where }I(z) &= \int_1^{+\infty} e^{z(t-1)} \frac{dt}{t^2}.
\end{align*}
By integration by parts ($e^{z(t-1)} = \frac{1}{z} \frac{d}{dt} e^{z(t-1)}$), we obtain
\begin{align*}
&I(z) = \frac{1}{z} ((0 - 1) + 2\int_1^{+\infty} \frac{e^{z(t-1)}}{t^3} dt),\; \text{hence}\\
&|I(z)| \leq \frac{1}{|z|} (1 + 2\int_1^{+\infty} \frac{dt}{t^3}) = \frac{2}{|z|},
\end{align*}
i.e.,
\begin{align*}
&|1 - z\, e^{-z} {\rm Ei}(z)| \leq \frac{2}{|z|}\quad \text{for $\Re z < 0$, and then}\\
&{\rm Ei}(z) \sim \frac{e^z}{z}\quad\text{when $|z| \rightarrow +\infty$, $\Re z < 0$},
\end{align*}
or, if $\Re z > 0$,
\begin{align*}
&|1 + z\, e^z {\rm Ei}(-z)| \leq \frac{2}{|z|}\quad \text{for $\Re z > 0$, and}\\
&{\rm Ei}(-z) \sim -\frac{e^{-z}}{z}\quad\text{when $|z| \rightarrow +\infty$, $\Re z > 0$}.
\end{align*}

\bibliography{Elastic_solution}

\begin{thebibliography}{22}
\bibitem [{Not()}]{Note}%
  \BibitemOpen
  \href@noop {} {}\bibinfo {note} {Note here that, as mentioned at the end of
  Sec.~\ref{GEN}, Young's equation Eq.~(\ref{Young}) is not valid (since
  $\gamma_{\rm bf} = \gamma_{\rm bf'}$ and $\varphi_{\rm f} > \pi/2$; see
  Fig.~\ref{Notations}b) and Eq.~(\ref{bff'equilibrium2b}) is valid, because
  (due to $a_{\rm r,\nu\nu} = 1$, $\gamma_{\rm bf} = \gamma_{\rm bf'}$, and
  $\sigma_{\rm bf',\tau\nu} = 0$) this equation reduces to $\cos\varphi_{\rm f}
  + \sin\varphi_{\rm f}\, \frac{\cos\varphi_{\rm b} + 1}{\sin\varphi_{\rm b}} =
  0$, i.e., $\sin\varphi_{\rm f} = -\sin(\varphi_{\rm b} + \varphi_{\rm f}) =
  \sin\varphi_{\rm f'}$, which is a consequence of $\varphi_{\rm f} =
  \varphi_{\rm f'}$}\BibitemShut {NoStop}%
\end{thebibliography}

\end{document}